\documentclass[10pt]{iopart}
\newcommand{\del}[2]{\frac{\partial #1}{\partial #2}}
\newcommand{\vect}[1]{\mbox{\boldmath$#1$}}
\usepackage{epsfig}
\begin{document}

\paper{Semiclassical Theory of Matter-Wave Detection}
\author{Nicholas K Whitlock, Stephen M Barnett and John Jeffers}
\address{Department of Physics, University of Strathclyde, Glasgow, G4 0NG, UK}
\ead{whitlock@phys.strath.ac.uk}

\begin{abstract}
We derive a semiclassical theory for the detection of
matter-waves. This theory draws on the theories of semiclassical
optical detection and of fluid mechanics. We observe that the
intrinsically dispersive nature of matter-waves is important in
deriving such a theory.
\end{abstract}

\nosections
In nature particles exist in either of two forms: those with
integer spin and those with half-integer spin. The former of these are
known as bosons, and the latter as fermions. At a microscopic level,
fermions are guided by the Pauli exclusion principle, which states
that no two fermions can occupy the same quantum state. On the other
hand, there is no such principle affecting bosons, and thus any number
can occupy the same state. This is the basis of Bose-Einstein
condensation, which was first proposed theoretically by Bose and
Einstein in 1924 \cite{bose:1924:1, einstein:1924:1,
einstein:1925:1}. A Bose-Einstein condensate (BEC) is a system in
which a macroscopic number of bosons occupy a single quantum
state. To achieve this with a dilute gas of atomic bosons requires
extremely low temperatures, such that the de-Broglie wavelength of the
particles become larger than their mean spacing. Hence no BEC was
experimentally realized in these systems until 1995
\cite{anderson:1995:1, davis:1995:1, bradley:1995:1} using bosonic
isotopes of Rb, Na and Li.

A BEC is created by cooling the atoms using optical and magnetic
forces, and then cooling them again using one of a number of
techniques. The BEC is then held in a magnetic trap, which is switched
off after a period of time to allow the atoms to expand so that imaging
can take place. A recent experiment by Robert {\it et al}
\cite{robert:2001:1} involved the creation of a BEC of metastable
triplet He (He$^*$) and highlighted the ability to count single atoms
falling from the trap after it was turned off. This allows for the
exciting possibility of more detailed investigation of the quantum
statistical properties of matter-waves.

As a first step into this field, we must model the detection of
matter-waves falling under gravity. We will use a simplistic model of
a BEC, not including the effects of interactions which exist between atoms
\cite{kagan:1996:1, castin:1996:1}. This will allow the features
specific to detection to be more readily illustrated. As we are considering
matter-waves, we can draw a direct analogy with the well known theory
of the detection of light waves, or photo-detection, which will be
outlined here. A more complete description can be found in many texts;
e.\,g. \cite{louden:book}.

The classical theory of photo-detection is based on the assumption
that the probability of an ionization event occuring in the
photo-detector in a time period $\rmd t$ is proportional to the
cycle-averaged intensity $\bar{I}(t)$ of the incoming light:
\begin{eqnarray}
p(t)\rmd t = \xi \bar{I}(t)\rmd t, \label{eq:pt}
\end{eqnarray}
where $\xi$ is a constant of proportionality which represents
the efficiency of the detector, including geometric factors such as
its area, and $\rmd t$ is sufficiently small that the probability of
more than one detection event occuring is negligible. In general, the
cycle-averaged intensity is taken to be
\begin{eqnarray}
\bar{I}(t) = \frac{1}{2}\epsilon_0c|E(t)|^2 \equiv cW(t), \label{eq:narrowb}
\end{eqnarray}
where $W(t)$ is the energy density. Under assumption (\ref{eq:pt}), if
we take a time interval from $t$ to $t+T$, then the probability of $m$
detection events occuring is
\begin{eqnarray}
P_m(T) =
\left\langle\frac{\bar{n}^m}{m!}\exp\left[-\bar{n}\right]\right\rangle,
\label{eq:pm}
\end{eqnarray}
where
\begin{eqnarray}
\bar{n} = \xi\int^{t+T}_t\rmd t'\bar{I}(t') \label{eq:photonbar}
\end{eqnarray}
and the angled brackets indicate a statistical average. From
(\ref{eq:pm}) we can evaluate the mean number of detection events to
be
\begin{eqnarray}
\langle m\rangle = \left\langle\bar{n}\right\rangle. \label{eq:avm}
\end{eqnarray}

We now wish to construct a semiclassical theory of matter-wave
detection by analogy with the theory of photo-detection presented
above. A natural way to proceed is to replace the electric field
$E(\vect{r},t)$ with the particle wavefunction
$\psi(\vect{r},t)$. Thus the matter-wave analogy to the expression for
$\bar{I}(t)$ in (\ref{eq:narrowb}) will be
$\left|\psi(\vect{r},t)\right|^2\bar{\vect{v}}$, where we have
included a characteristic velocity $\bar{\vect{v}}$. This is
in direct analogy with the velocity of light $c$ in the
photo-detection theory and is of vectorial nature to allow for
matter-waves which are not travelling perpendicular to
the detector. It is also required so that the equations have the
correct dimensionality. In the analysis that follows, $\bar{\vect{v}}$ will
be associated with the mean velocity of the wavepacket. The
probability of detection over a time interval from
$t$ to $t+T$ would again be given by (\ref{eq:pm}) and the average
number of counts $\langle m\rangle$ by (\ref{eq:avm}) where instead of
(\ref{eq:photonbar}), we have
\begin{eqnarray}
\bar{n} = \xi\int^{t+T}_t\rmd t'\int_A
\left|\psi(\vect{r},t')\right|^2\bar{\vect{v}}\cdot\rmd\vect{A}.
\label{eq:wrongnbar}
\end{eqnarray}
We have now explicitly included the area of the detector $A$, and
$\rmd\vect{A}$ is the infinitesimal area element normal to the surface
of the detector. If we assume that the particle wavefunction is
normalized so that it contains on average $N$ particles, then for all
times $t$
\begin{eqnarray}
\int_{-\infty}^{\infty}\rmd^3\vect{r}\left|\psi(\vect{r},t)\right|^2 = N.
\end{eqnarray}
If the detector is of perfect efficiency then we would expect that for a
wavepacket falling under gravity, a sufficiently long detection window
and large detection area would produce a mean of $N$ detection
events. This means that from (\ref{eq:avm}) we might expect that as
$T\rightarrow\infty$,
\begin{eqnarray}
\xi\int_{-\infty}^{\infty}\rmd t\int_A
\left|\psi(\vect{r},t)\right|^2\bar{\vect{v}}\cdot\rmd\vect{A} = N,
\label{eq:wrongint}
\end{eqnarray}
for the value of $\xi$ corresponding to a perfectly efficient detector.

By drawing analogy with photo-detection of light waves, we have
derived (\ref{eq:wrongnbar}) which includes the characteristic velocity
$\bar{\vect{v}}$. As a first approximation we might expect that this
will be the mean velocity of the wavepacket. This is not an
approximation for light in free space because free space is not
dispersive; at all frequencies light travels at $c$. For matter-waves,
however, free space {\it is} dispersive.

In order to take into account matter-wave dispersion we ought to base
our theory of detection on the flux-density of particles: the mean
rate at which particles cross a unit area of the detector. As particle
number is a conserved quantity it must satisfy an equation of
continuity \cite{landau:1941:1,khalatnikov:book}
\begin{eqnarray}
\del{}{t}\left|\psi(\vect{r},t)\right|^2 +
\nabla\cdot\vect{J}(\vect{r},t) = 0, \label{eq:fluidpart}
\end{eqnarray}
where $\vect{J}$ is the particle flux-density. This equation is of the
same form as the one for local charge conservation in electromagnetic
theory or, more relevantly for our purpose, relating particle density
$\rho$ and particle flux-density $\vect{J} = \rho\vect{v}$ in fluid
mechanics \cite{landlif:v6}. From (\ref{eq:fluidpart}) we obtain a
particle flux-density of the form
\begin{eqnarray}
\vect{J}(\vect{r},t) =
\frac{\hbar}{m}\mathrm{Im}\left\{\psi^*(\vect{r},t)
\nabla\psi(\vect{r},t)\right\}. \label{eq:flux}
\end{eqnarray}
As this is analogous to the particle flux-density $\vect{J} =
\rho\vect{v}$ from fluid mechanics, it seems reasonable that
(\ref{eq:wrongnbar}) would become
\begin{eqnarray}
\bar{n} = \xi\int_t^{t+T}\rmd
t'\int_A\vect{J}(\vect{r},t')\cdot\rmd\vect{A}. \label{eq:correctnbar}
\end{eqnarray}
Indeed, in electromagnetic theory a relation similar to
(\ref{eq:fluidpart}) exists between the energy density $W(t)$ and the
Poynting vector (which gives the energy flux-density)
\cite{poynting:1884:1,jackson:book}. \Eref{eq:narrowb} is thus an
approximation which holds in most experimentally realizable
situations. In situations where this approximation is invalid, the
cycle-averaged intensity in (\ref{eq:narrowb}) must be replaced by the
magnitude of the Poynting vector.

In order to illustrate fully the difference between the theories given
by (\ref{eq:wrongnbar}) and (\ref{eq:correctnbar}), it is instructive to
evaluate both expressions in the case of the detection of a wavepacket
falling in the $z$-direction under gravity onto a flat, large-area
detector aligned parallel to the $x$-$y$ plane. Such a system closely
models the He$^*$ experiment mentioned at the beginning of this
paper, and it is one in which we would expect all particles to fall
onto the detector, which will allow us to check the expression for
$\langle m\rangle$.

In evaluating the probability of detection for a wavepacket falling
under gravity we will need to calculate the form of the
matter-wave. We consider a model BEC, released at time $t=0$,
described by a Gaussian wavefunction centred at $\vect{r}_0$ with width
parameter $w$,
\begin{eqnarray}
\psi(\vect{r},0) = N^{\frac{1}{2}}(\pi
w^2)^{-\frac{3}{4}}\exp\left\{-\frac{|\vect{r}-\vect{r}_0|^2}{2w^2}\right\}.
\end{eqnarray}
The standard solution to the Schr\"{o}dinger equation takes the form
\begin{eqnarray}
\psi(\vect{r},t) = \exp\left\{\frac{-\rmi t\hat{H}}{\hbar}\right\}
\psi(\vect{r},0), \label{eq:wavet}
\end{eqnarray}
where $\hat{H}$ is the Hamiltonian, which in this case has the standard
kinetic energy term and a gravitational potential term
\begin{eqnarray}
\hat{H} = \frac{\hat{p}^2}{2m} + mg\hat{z}.
\end{eqnarray}
We have taken the zero of gravitational potential energy to be at
$z=0$. When written in the position representation, this becomes
\begin{eqnarray}
\hat{H} = \frac{-\hbar^2}{2m}\nabla^2 + mgz,
\end{eqnarray}
and so the particle wavefunction at a later time $t$ will be
\begin{eqnarray}
\fl\psi(\vect{r},t) &=& N^{\frac{1}{2}}(\pi w^2)^{-\frac{3}{4}}
\exp\left\{-\frac{\rmi t}{\hbar}\left(\frac{-\hbar^2}{2m}\nabla^2 +
mgz\right)\right\}
\exp\left\{-\frac{|\vect{r}-\vect{r}_0|^2}{2w^2}\right\}.
\end{eqnarray}
Techniques outlined in \cite{barnett:book} allow us to evaluate the
deriviatives in this expression and obtain a wavefunction for the
matter-wave falling under gravity
\begin{eqnarray}
\fl\psi(\vect{r},t) &=& N^{\frac{1}{2}}\pi^{-\frac{3}{4}}
\left(\frac{w}{w^2+\rmi t\frac{\hbar}{m}}\right)^{\frac{3}{2}}
\exp\left\{-\frac{\left|\vect{r}
-\vect{R}(t)\right|^2}{2\left(w^2+\rmi t\frac{\hbar}{m}\right)}\right\}
\exp\left\{-\frac{\rmi tmg}{\hbar}\left(z+\frac{1}{6}gt^2\right)\right\},
\label{eq:zwave}
\end{eqnarray}
where we have defined the average ``classical'' position of the
particle $\vect{R}(t) = \langle\vect{r}(t)\rangle =
\vect{r}_0-\frac{1}{2}gt^2\hat{\vect{k}}$, which gives the position of
the centre of the wavepacket. As the wavepacket is accelerating from
rest under gravity, the integral in (\ref{eq:wrongnbar}) will be given
by
\begin{eqnarray}
\int_A\left|\psi(\vect{r},t)\right|^2\bar{\vect{v}}\cdot\rmd\vect{A} =
gt\int\!\!\!\int\left|\psi(\vect{r},t)\right|^2\rmd x\rmd
y. \label{eq:wrongdp}
\end{eqnarray}
It can be seen that the expression in (\ref{eq:wrongdp}) depends on
$\exp\{-(z-z_0)^2\}$ and thus depends on the height that the
wavepacket starts above the detection screen. With this taken into
account, one can see that the integral of (\ref{eq:wrongdp}) over all
time cannot give a constant value of $N$, and so the expression in
(\ref{eq:wrongint}) cannot hold for any $\xi$ which is solely
dependent on detector properties. This result can be verified
numerically.

If we now use (\ref{eq:flux}) to calculate the flux-density of
particles for this system, we obtain an expression for the integral in
(\ref{eq:correctnbar})
\begin{eqnarray}
\int_A\vect{J}(\vect{r},t)\cdot\rmd\vect{A} = \left[gt -
\frac{z-z_0+gt^2/2}{t+w^4m^2/(\hbar^2t)}\right]
\int\!\!\!\int\left|\psi(\vect{r},t)\right|^2\rmd x\rmd y.
\label{eq:correctdp}
\end{eqnarray}
It is straightforward to show that the integral of this expression
over all time gives the average number of particles in the wavepacket
$N$. Thus from (\ref{eq:correctnbar}) we can see that the constant of
proportionality $\xi$ is in fact the efficiency of the detector
$\eta$, which takes values between 0 and 1.

\begin{figure}
\begin{center}
\epsfig{figure=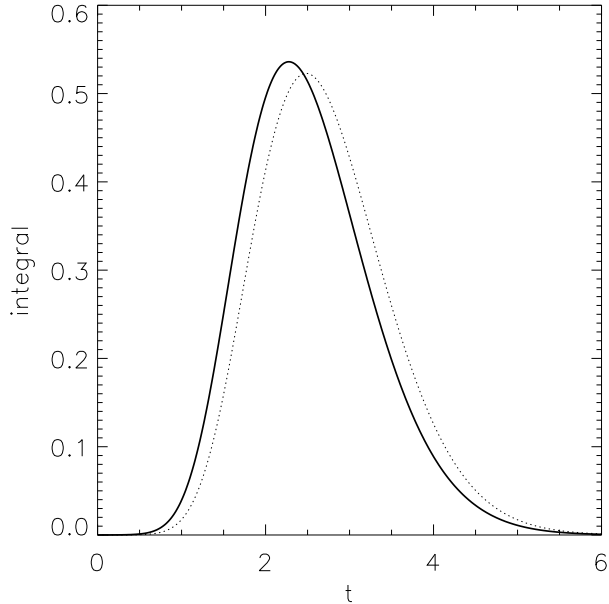,width=80mm}
\end{center}
\caption{\label{fig:compar}Comparison of integrals in the intensity
term for the correct and incorrect detection theories, as a
function of time. The units have been chosen in such a way that
$g=\hbar/m=1$, and we have chosen $w=1$ and $z_0=3$. The solid line
shows the correct theory (\ref{eq:correctdp}) and the dotted
line shows the incorrect theory (\ref{eq:wrongdp}).}
\end{figure}

It is clear to see that the expression obtained in
(\ref{eq:correctdp}) is that from (\ref{eq:wrongdp}) plus an
additional correction, which is a height-dependent velocity term. This
additional velocity term is a direct consequence of the dispersive
nature of free space for matter-waves. From (\ref{eq:zwave}) it is
clear that the wave undergoes dispersion as it falls under
gravity. The detection theory based on (\ref{eq:wrongdp}) assumes that
this dispersed wavepacket propagates through the detection plane at
the mean packet velocity. The detection formula in
(\ref{eq:correctdp}) based on particle flux does not make this
assumption and the factor $\hbar/m$ which quantifies the dispersion of the
wave in (\ref{eq:zwave}) also appears in the detection formula. If
this factor is taken to zero either by taking $\hbar\rightarrow0$ or
$m\rightarrow\infty$, then the dispersion in (\ref{eq:zwave})
disappears, as does the additional velocity term in
(\ref{eq:correctdp}). The time variation of the integrals given by the
two different theories are plotted in \fref{fig:compar}, where it can
be seen that the differences in the expressions are quite
pronounced: in a detection theory which takes account of dispersion
the majority of particles will arrive earlier than they would in a
detection theory in which dispersion is not correctly accounted for.

From (\ref{eq:zwave}) we can see that if the factor $\hbar t/(mw^2)$
is greater than unity, the wavepacket becomes significantly wider (due
to dispersion), and this dispersion ought to be taken account of in
detection. As
an example of how important dispersion is in the
system under consideration, we take values from the He$^*$ experiment
presented in \cite{robert:2001:1}. The time of flight of atoms here is
0.1s and the mass of a He$^*$ atom is $6.68\times10^{-27}$ kg. We thus
find that the dispersion factor will be important for any wavepacket
with an initial width of less than 0.1mm.

We have described in this paper the construction of a semiclassical
theory of matter-wave detection, drawing on the well known theory of
photo-detection. It is the intrinsically dispersive nature of
matter-waves which prevents the direct analogy from working. We must
instead consider the flux-density of particles, which gives an
additional velocity term. Indeed if light passes through and is
detected in a dispersive medium, the magnitude of the Poynting vector,
which represents the flux-density of energy, must be used in place of
(\ref{eq:narrowb}).

An instructive ``next step'' will be to consider the second quantized
version of this theory. We would expect that in doing this, a
situation of no detection in the early part of the wavepacket would
feed back to modify the later part of the wavepacket. It is
also clear that quantities other than particle number - such as
energy, momentum and angular momentum - can be conserved. We intend
to investigate these conservation laws, fluxes and the deposition of
such quantities on a detection screen.

\ack
We are grateful to Alain Aspect for first suggesting that we consider
this problem and to James Cresser and Erika Andersson for helpful
discussions. The authors would like to thank the Overseas Research
Students Awards Scheme, the University of Strathclyde, the Royal
Society of Edinburgh, the Scottish Executive Education and Lifelong
Learning Department and the European Commision (project QUANTIM
(IST-2000-26019)) for financial support.


\end{document}